\documentclass[prl,reprint,nofootinbib,superscriptaddress,preprintnumbers,showpacs]{revtex4-1}
\usepackage{amsfonts}
\usepackage{amssymb}
\usepackage{amsmath}
\usepackage{graphicx,color}
\usepackage{dcolumn}
\usepackage{epsfig}
\usepackage{bm}
\usepackage{bbm}
\usepackage{ulem}
\usepackage{hyperref}
\usepackage{lineno}

\usepackage{color}

\def\gtap{\ \raise.3ex\hbox{$>$\kern-.75em\lower1ex\hbox{$\sim$}}\ }
\def\ltap{\ \raise.3ex\hbox{$<$\kern-.75em\lower1ex\hbox{$\sim$}}\ }
\begin{document}

\title{
Triangle singularities in $\bar{B}^0\to \chi_{c1}K^-\pi^+$
relevant to $Z_1(4050)$ and $Z_2(4250)$
}
\author{Satoshi X. Nakamura}
\email{satoshi@ustc.edu.cn}
\affiliation{
University of Science and Technology of China, Hefei 230026, 
People's Republic of China
}
\affiliation{
State Key Laboratory of Particle Detection and Electronics (IHEP-USTC), Hefei 230036, People's Republic of China}

\begin{abstract}
 $Z_1(4050)$ and $Z_2(4250)$
observed in $\bar{B}^0\to\chi_{c1}K^-\pi^+$ by the Belle Collaboration
are candidates of 
charged charmonium-like states that minimally include two quarks and
two antiquarks.
 While $Z_1(4050)$ and $Z_2(4250)$ have been interpreted as tetraquark
 states previously,
 we propose a completely different scenario based on
a kinematical effect called the triangle singularity.
 We demonstrate that the triangle singularities cause
 in the $\chi_{c1}\pi^+$ invariant mass distribution
 resonance-like bumps
that fit very well the Belle data.
 If these bumps are simulated by the
 $Z_1(4050)$ and $Z_2(4250)$ resonance excitations,
 the spin-parity of them
 are predicted to be $1^-$ for $Z_1(4050)$ and $1^+$ or $1^-$ for $Z_2(4250)$.
The bump corresponding to $Z_1(4050)$ has a highly asymmetric shape, 
which the Belle data exactly indicate.
 We show that the asymmetric shape originates from
 an interplay between the triangle singularity and 
the opening of the $X(3872)\pi^+$ channel near 
the triangle-singularity energy.
 This characteristic lineshape could be used to discriminate different
 interpretations of $Z_1(4050)$.
 An interesting prediction from interpreting 
 $Z_1(4050)$ and $Z_2(4250)$ as the triangle singularities is that
 similar bumps caused by the same mechanisms possibly appear also
 in $\bar{B}^0\to J/\psi K^-\pi^+$ data;
 the already observed $Z_c(4200)$ corresponds to $Z_2(4250)$ of $J^P=1^+$.
\end{abstract}

\maketitle

\section{Introduction}

$Z_1(4050)$ and $Z_2(4250)$
($X(4050)$ and $X(4250)$ in the Particle Data Group (PDG) notation~\cite{pdg})
 were observed in the Belle experiment
as resonance-like structures in the  
$\chi_{c1}\pi^+$ invariant mass distribution
of $\bar B^0\to\chi_{c1} K^-\pi^+$~\cite{belle_z4050}
\footnote{
We implicitly include the charge conjugate mode
throughout. }.
It was not possible to determine the spin($J$)-parity($P$) of these states.
The following analysis in the BaBar experiment~\cite{babar_z4050}
did not confirm them because the resonance-like signals were only barely discernible
and insignificant.
$Z_1(4050)$ and $Z_2(4250)$ are clearly candidates of
charged charmonium-like states that minimally include four quarks and
 thus not belonging to the conventional quark model picture.
 In order to understand the QCD dynamics and its consequence in the
 non-perturbative regime,
 it is highly desirable to establish their existence with higher
 statistics data in the experimental side, 
 and to clarify their identities such as tetraquark,
 hadron molecule, or kinematical effect in the theoretical side.

 Previous theoretical interpretations of $Z_1(4050)$ and $Z_2(4250)$
 are mainly categorized into tetraquark and hadron-molecule.
Within the tetraquark picture:
(1) a diquark-antidiquark state is [not] assigned to 
$Z_2(4250)$ [$Z_1(4050)$]~\cite{z4050_diquark1};
(2) $Z_1(4050)$ is described by a
 molecular-like tetraquark picture~\cite{z4050_diquark2};
(3) $Z_1(4050)$ and $Z_2(4250)$ are described with 
 tetraquarks based on a color flux-tube model~\cite{z4050_diquark3};
 (4) $J^{PC}=0^{++}$ diquark-antidiquark state is not assigned to 
$Z_1(4050)$ and $Z_2(4250)$ using QCD sum rule (QCDSR)~\cite{z4050_qcdsr1};
 (5) a tetraquark state is assigned to 
$Z_2(4250)$ using QCDSR~\cite{z4050_qcdsr2}.
Meanwhile, within the hadron-molecule picture:
(1) meson-exchange models disfavor hadron-molecule pictures for 
$Z_1(4050)$ and $Z_2(4250)$~\cite{z4050_molecule1,z4050_molecule2,z4050_molecule3};
(2) $D_1\bar{D}$ molecule state is
assigned to $Z_2(4250)$ based on QCDSR~\cite{z4050_qcdsr4}.
For a more complete summary, 
see reviews~\cite{review_chen,review_raphael}.

\begin{figure*}[t]
\begin{center}
\includegraphics[width=1\textwidth]{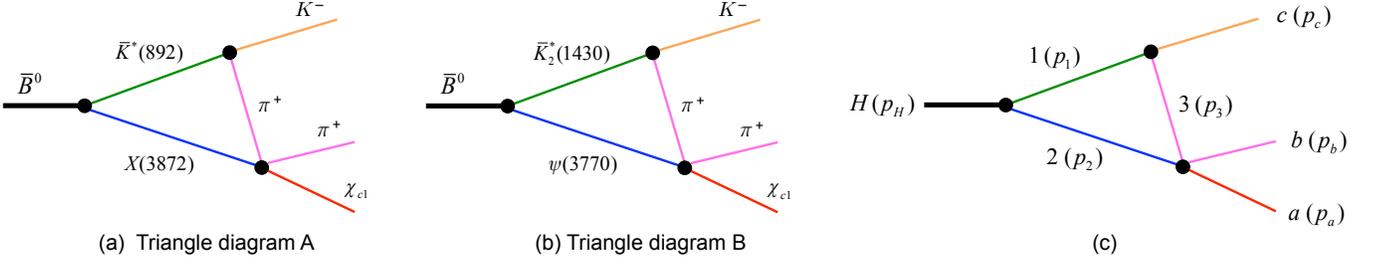}
\end{center}
 \caption{
Triangle diagrams contributing to 
 $\bar B^0\to\chi_{c1} K^-\pi^+$ (a,b).
The triangle singularity from
the diagram (a) [(b)] generates a
$Z_1(4050)$ [$Z_2(4250)$]-like bump
in the $\chi_{c1}\pi^+$
invariant mass distribution. 
 A generic triangle diagram (c) defines particle labels and,
 in the parentheses, their momenta.
 }
\label{fig:diag}
\end{figure*}

In this work, we propose a completely different
interpretation of $Z_1(4050)$ and $Z_2(4250)$.
This is to associate $Z_1(4050)$ and $Z_2(4250)$ with
triangle singularities (TS)~\cite{landau,coleman,s-matrix},
which is a kinematical effect,
arising from triangle diagrams
depicted in Figs.~\ref{fig:diag}(a) and \ref{fig:diag}(b)
(we refer to them as the triangle diagrams A and B hereafter),
respectively.
The diagrams consist of experimentally well-established hadrons
including $X(3872)$ ($\chi_{c1}(3872)$ in the PDG).
The TS can occur only when 
three particles in the loop
go through a classically allowed kinematics
(on-shell and collinear
in the center-of-mass (CM) frame)
 at the same time,
 and can generate a resonance-like spectrum bump;
 see an illustrative discussion in Ref.~\cite{TS-Pc2}
 for a mathematical detail.
 Applications of TS to phenomenology have become popular these
 days~\cite{wu1,wu2,TS-Pc,TS-Pc3,TS-Pc2,TS-a1,TS-a1-2,ts1,ts2,ts3,ts4,ts5,ts6},
 such as 
 explaining
 isospin violations in $\eta(1405/1475)\to 3\pi$~\cite{wu1,wu2}, 
and interpreting
recently discovered
hidden charm pentaquark $P_c(4450)^+$~\cite{TS-Pc,TS-Pc3,TS-Pc2}
and $a_1(1420)$~\cite{TS-a1,TS-a1-2}.

 Recently we also applied TS~\cite{ts_zc4430} to interpreting 
  $Z_c(4430)$~\cite{belle_z4430_2008,belle_z4430,lhcb_z4430}
 and $Z_c(4200)$~\cite{belle_z4200},
  charged charmonium-like state candidates,
 observed in
  $\bar B^0\to\psi(2S) K^-\pi^+$ and $J/\psi K^-\pi^+$.
We successfully explained their properties ($J^P$, mass, width,
 Argand plot) extracted in the experiments.
 The presence [absence] of $Z_c(4200)$[$Z_c(4430)$]-like contribution in 
 $\Lambda_b^0\to J/\psi p\pi^-$~\cite{lhcb_z4200_Lb}
 was also explained in terms of the TS.
 The present work shows that $Z_1(4050)$ and $Z_2(4250)$ can also be
 consistently interpreted as TS, provided 
 the TS have experimentally detectable strengths.
We demonstrate that the triangle diagram A [B]
creates a $Z_1(4050)$ [$Z_2(4250)$]-like bump in 
the $\chi_{c1}\pi^+$ invariant mass ($m_{\chi_{c1}\pi}$) distribution
of $\bar B^0\to\chi_{c1}K^-\pi^+$.
Simulating the bumps with the $Z_1(4050)$ and $Z_2(4250)$
resonance excitations, 
$J^P=1^-$ and $1^\pm$ are predicted, respectively. 
The Breit-Wigner masses and widths fitted to the bumps
agree very well with 
those of $Z_1(4050)$ and $Z_1(4250)$ from the Belle analysis~\cite{belle_z4050}.
The $Z_1(4050)$-like bump has a highly asymmetric shape as
the Belle data exactly indicates.
We clarify that the opening of the $X(3872)\pi^+$ channel 
near the TS energy of $m_{\chi_{c1}\pi}\sim 4.02$~GeV
is responsible for it. 
This characteristic bump shape could discriminate
different interpretations of $Z_1(4050)$.
We point out that
the triangle singularities for $Z_1(4050)$ and $Z_2(4250)$
could also generate similar bumps in $\bar{B}^0\to J/\psi K^-\pi^+$;
the already observed $Z_c(4200)$ corresponds to $Z_2(4250)$ of $J^P=1^+$.

\section{model}

We calculate the $\bar{B}^0\to \chi_{c1}K^-\pi^+$
decay amplitudes due to the triangle diagrams A and B
of Fig.~\ref{fig:diag}.
A general formula for the decay amplitude is given by 
\begin{eqnarray}
 T_{abc,H} &=& \int d\bm{p}_1\,
  { v_{ab;23}(\bm{p}_a,\bm{p}_b;\bm{p}_2,\bm{p}_3)\,
  \Gamma_{3c,1}(\bm{p}_3,\bm{p}_c;\bm{p}_1)
  \over
  E - E_2(\bm{p}_2) - E_3(\bm{p}_3) - E_c(\bm{p}_c) + i\epsilon
  }
  \nonumber \\
  &\times&  { 1  \over
  E - E_1(\bm{p}_1) - E_2(\bm{p}_2) }
  \Gamma_{12,H}(\bm{p}_1,\bm{p}_2;\bm{p}_H)
  \ ,
  \label{eq:amp}
\end{eqnarray}
where we have used the
particle labels and their momenta in 
Fig.~\ref{fig:diag}(c).
Spin states of the intermediate particles are implicitly summed. 
The total energy in the CM frame is denoted by $E$, 
while the energy of a particle $x$ is 
$E_x(\bm{p}_x)=\sqrt{\bm{p}^2_x+m^2_x} - i\Gamma_x/2$
with the mass $m_x$, momentum $\bm{p}_x$, and
width $\Gamma_x$; $\Gamma_x\neq 0$ only for 
unstable intermediate particles 1 and 2.
We use the mass and width values of the PDG average~\cite{pdg}.
Because $X(3872)$ has a very small width
($\Gamma_{(X(3872)}<1.2$~MeV),
we set it to zero in calculations.

The pion-charmonium interaction is denoted by $v_{ab;23}$ in Eq.~(\ref{eq:amp}).
The particles 2
is either $X(3872)$[$J^P=1^+$] or $\psi(3770)[1^-]$, the particle $a$ is $\chi_{c1}[1^+]$, and 
the particles 3 and $b$ are pions$[0^-]$.
In calculating the triangle diagram A, 
we use an $s$-wave interaction:
\begin{eqnarray}
 v^{({\rm A})}_{ab;23}(\bm{p}_a,\bm{p}_b;\bm{p}_2,\bm{p}_3)
  = f^{01}_{ab}(p_{ab}) f^{01}_{23}(p_{23})\,
  \bm{\epsilon}^*_a\cdot\bm{\epsilon}_2 \ ,
\label{eq:contact}
\end{eqnarray}
where polarization vectors for 
the particles $a$ and 2 are denoted by
 $\bm{\epsilon}_a$ and $\bm{\epsilon}_2$, respectively.
The quantities
$f^{01}_{ab}(p_{ab})$ and $f^{01}_{23}(p_{23})$
are form factors that 
will be defined in Eq.~(\ref{eq:ff});
the momentum of the particle $i$ in the $ij$-CM frame is denoted by
$\bm{p}_{ij}$ and $p_{ij}=|\bm{p}_{ij}|$.
This interaction leaves
an $s$-wave $\chi_{c1}\pi^+$ pair in the final state.
Therefore, if a spectrum bump is created by 
the triangle diagram A
in the $m_{\chi_{c1}\pi}$ distribution
and is simulated by a resonance-excitation, 
the resonance has $J^P=1^-$.

Regarding $v_{ab;23}$ for the triangle diagram B,
where the intrinsic parity is different between the incoming and
outgoing states, we use
\begin{eqnarray}
 v^{({\rm B})}_{ab;23}(\bm{p}_a,\bm{p}_b;\bm{p}_2,\bm{p}_3)
  = f^{11}_{ab}(p_{ab}) f^{01}_{23}(p_{23})\,
  \bm{\epsilon}^*_2\!\!\cdot\!
\bm{\epsilon}_a\!\times\!\bm{p}_{ab},
\label{eq:contact2}
\end{eqnarray}
which converts $s$-wave $\psi(3770)\pi^+$ into
$p$-wave $\chi_{c1}\pi^+$.
 A resonance that simulates 
a $\chi_{c1}\pi^+$ spectrum bump from
the triangle diagram B has $J^P=1^+$.
In Eqs.~(\ref{eq:contact}) and (\ref{eq:contact2}),
the incoming 23-pair is in $s$-wave
and can create a sharp TS bump,
being free from the centrifugal barrier.
For the triangle diagram B, however, we also examine an
interaction of $p$-wave $\psi(3770)\pi^+$ going to 
$s$-wave $\chi_{c1}\pi^+$ because 
the $\psi(3770)\pi^+$ threshold is rather below the TS energy ($\sim 4.25$~GeV)
and the centrifugal barrier would not be so effective.
Such an interaction is
\begin{eqnarray}
 v^{({\rm B'})}_{ab;23}(\bm{p}_a,\bm{p}_b;\bm{p}_2,\bm{p}_3)
  = f^{01}_{ab}(p_{ab}) f^{11}_{23}(p_{23})\,
  \bm{\epsilon}^*_2\!\!\cdot\!
\bm{\epsilon}_a\!\times\!\bm{p}_{23},
\label{eq:contact3}
\end{eqnarray}
and the $\chi_{c1}\pi^+$ pair seems to be from 
a $J^P=1^-$ resonance.

\begin{figure*}[t]
\begin{center}
\includegraphics[width=1\textwidth]{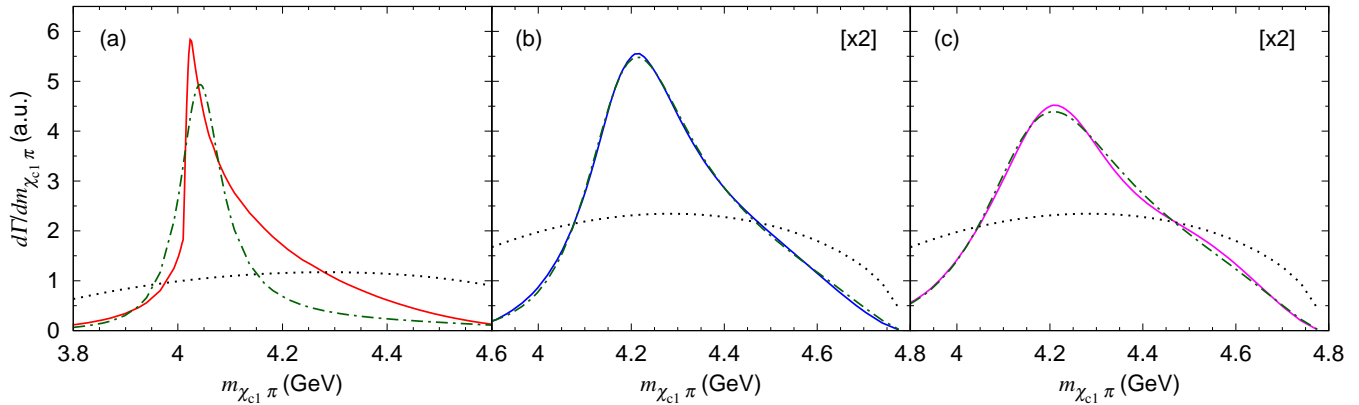}
\end{center}
\caption{
$\chi_{c1}\pi^+$ invariant mass distributions for 
 $\bar B^0\to\chi_{c1} K^-\pi^+$.
The red (blue and magenta) solid curve in the panel (a) [(b) and (c)] is obtained from
the triangle diagram A [B].
The interactions $v_{ab;23}$ of Eqs.~(\ref{eq:contact}), (\ref{eq:contact2}),
and (\ref{eq:contact3}) are used for (a), (b), and (c), respectively.
 The green dash-dotted curves are from Breit-Wigner amplitudes fitted
to the solid curves.
The dotted curves are the phase-space distributions.
The solid and dotted curves are 
normalized to give unity when integrated with respect to
 $m_{\chi_{c1}\pi}$.
 The scale for the panels (b) and (c) has been doubled.
}
\label{fig:spec}
\end{figure*}
The $R\to ij$ vertex function, $\Gamma_{ij,R}$
in Eq.~(\ref{eq:amp}), 
is given by
\begin{eqnarray}
  \Gamma_{ij,R}(\bm{p}_i,\bm{p}_j;\bm{p}_R)
   &=& \sum_{LS}f^{LS}_{ij}(p_{ij}) (s_is_i^zs_js_j^z|SS^z)
   \nonumber \\
&\times&   (LM S S^z|S_RS^z_R)
   Y_{LM} (\hat{p}_{ij}) \ ,
  \label{eq:vertex}
\end{eqnarray}
with $Y_{LM}$ being spherical harmonics. 
Clebsch-Gordan coefficients are
$(abcd|ef)$ in which
the spin of a particle $x$ is 
denoted by $s_x$ and its $z$-component $s_x^z$.
We use the form factor $f^{LS}_{ij}(p_{ij})$ in the form of
\begin{eqnarray}
 f^{LS}_{ij}(p) =
g^{LS}_{ij} {p^L\over \sqrt{E_i(p) E_j(p)}}
\left(\frac{\Lambda^2}{\Lambda^2+p^2}\right)^{2+(L/2)}\ ,
\label{eq:ff}
\end{eqnarray}
which is parametrized with a coupling $g^{LS}_{ij}$ and
a cutoff $\Lambda$.
For the $1\to 3c$ and $23\to ab$ interactions,
a nonzero value of $g^{LS}_{ij}$ is allowed for
only one set of $\{L,S\}$.
While the actual values of $g^{LS}_{ij}$ 
for the $1\to 3c$ processes
can be determined using the 
$K^*(892)$ and $K^*_2(1430)$ decay widths,
experimental and Lattice QCD inputs are currently missing to determine the
couplings for the $23\to ab$ interactions.
Experimentally, 
$X(3872)\to \chi_{c1}\pi^+\pi^-$ has not yet been seen~\cite{x3872_decay},
perhaps because 
$X(3872)$ has a very small width and the phase-space for this final
state is small;
$\psi(3770)\to \chi_{c1}\pi\pi$ is not kinematically allowed.
Here we assume that these couplings are strong enough and set them arbitrary.

Regarding the weak vertices for the $H\to 12$ decays,
$g^{LS}_{ij}\neq 0$ is allowed for 
several sets of $\{L,S\}$ but their values are
currently difficult to estimate due to the lack of data. 
However, the details of these vertices would not be crucial in this work
because the main conclusions
are essentially determined by the kinematical effects
once the structure of $v_{ab;23}$ is fixed as
Eqs.~(\ref{eq:contact})-(\ref{eq:contact3}).
Thus we assume simple structures and detectable strengths.
We set $g^{LS}_{ij}\neq 0$ only for $S=|s_1-s_2|$ 
(exception: $S=2$ when using Eq.~(\ref{eq:contact3}))
and the lowest allowed $L$;
$g^{LS}_{ij}=0$ for the other $\{L,S\}$.
We use the cutoff $\Lambda=1$~GeV in Eq.~(\ref{eq:ff})
throughout unless otherwise stated.

The interactions of Eqs.~(\ref{eq:contact})-(\ref{eq:vertex}),
evaluated in the CM frame of the two-body subsystem,
are further multiplied by kinematical factors to account for the Lorentz
transformation to the total three-body CM frame; see
Appendix~C of Ref.~\cite{3pi}.
The Dalitz plot distribution for $H\to abc$ 
is calculated with $T_{abc,H}$
 of Eq.~(\ref{eq:amp})
following the procedure detailed in 
Appendix~B of Ref.~\cite{3pi}.

\section{results}

In Fig.~\ref{fig:spec},
we present the $\chi_{c1}\pi^+$
invariant mass distributions
for $\bar B^0\to\chi_{c1} K^-\pi^+$.
The triangle diagram A [B]
gives the red [blue and magenta] solid curve in Fig.~\ref{fig:spec}(a)
[\ref{fig:spec}(b) and \ref{fig:spec}(c)].
We also show the phase-space distributions (black dotted curves).
The triangle singularity creates 
clear resonance-like peaks at 
$m_{\chi_{c1}\pi}\sim 4.02$~GeV in panel (a) and
$m_{\chi_{c1}\pi}\sim 4.22$~GeV in panels (b) and (c).
It is interesting to observe in Fig.~\ref{fig:spec}(a)
that the bump
has a significantly asymmetric shape.

\begin{figure}[b]
\begin{center}
\includegraphics[width=.5\textwidth]{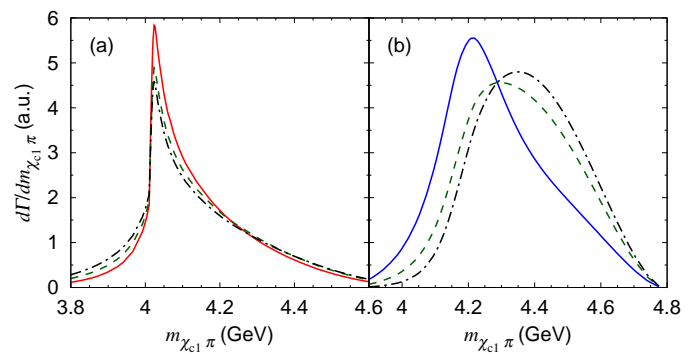}
\end{center}
 \caption{
 Cutoff ($\Lambda$) dependence of the spectrum shapes
generated by triangle diagrams.
 The panels (a) and (b) correspond to
 the triangle diagrams A and B, respectively.
 The red and blue solid curves are the same as those in
 Fig.~\ref{fig:spec}(a) and \ref{fig:spec}(b), respectively,
 where $\Lambda=1$~GeV.
 The green dashed, and black dash-dotted curves are
 obtained with $\Lambda=$ 1.5, and 2~GeV,  respectively.
 All the curves are normalized as in Fig.~\ref{fig:spec}.
 }
\label{fig:cutoff}
\end{figure}
We examine how the spectrum shapes shown in Fig.~\ref{fig:spec}
depend on the cutoff $\Lambda$
of the form factor in Eq.~(\ref{eq:ff}).
In Fig.~\ref{fig:cutoff},
the spectrum shapes calculated with
$\Lambda=$ 1, 1.5, and 2~GeV are shown.
The resonance-like peak structures
due to the kinematical singularities
are clearly stable over the
reasonable cutoff range.
In particular, the peak positions of the spectra from the triangle
diagram A are, as in Fig.~\ref{fig:cutoff}(a), 
almost the same; the width is somewhat broadened as $\Lambda$ increases.
The peak position and width of the spectrum from the triangle diagram B are more
dependent on $\Lambda$.
This would be related to the fact that the unstable particles in
the triangle diagram B have wider widths 
than those in the triangle diagram A,
thereby pushing the exact TS further away from the physical region.
Now we simulate the spectra
using the conventional resonance-excitation mechanisms,
and determine the masses and widths of the fake resonances. 
The Dalitz plot distributions from the triangle diagrams A and B
 are fitted with the mechanism of
$\bar B^0\to Z K^-$ followed by $Z \to\chi_{c1}\pi^+$.
The Breit-Wigner form of Ref.~\cite{belle_z4430} is used for 
the $Z$ propagation.
This $Z$-excitation mechanism includes fitting parameters such as 
the Breit-Wigner mass, width, and also the cutoff in
Eq.~(\ref{eq:ff}) that describes the vertices. 
The kinematical region included in the fit covers
the Dalitz plot distribution larger than 10\% of the peak height. 
The obtained fits are shown by the green
dash-dotted curves in Fig.~\ref{fig:spec}.
The highly asymmetric bump, the red solid curve in 
Fig.~\ref{fig:spec}(a), is not well fitted with the Breit-Wigner form,
while the bumps in Figs.~\ref{fig:spec}(b) and \ref{fig:spec}(c) are reasonably fitted.
\begin{table}[b]
 \caption{\label{tab:BW_param}
 Spin-parity $J^P$ (third row),
 Breit-Wigner mass in MeV (fourth row), and
 width in MeV (fifth row) for $Z_1(4050)$ and $Z_2(4250)$.
The Breit-Wigner parameters for 
 $Z_1(4050)$ [$Z_2(4250)$]
are extracted by fitting the Dalitz plot distributions for 
$\bar B^0\to\chi_{c1} K^-\pi^+$
generated by triangle diagram of Fig.~\ref{fig:diag}(a) [\ref{fig:diag}(b)].
The parameter ranges are from the cutoff dependence.
 The parameters from the Belle analysis~\cite{belle_z4050}
 are also shown; 
the first (second) errors are statistical (systematic). 
}
\begin{ruledtabular}
\renewcommand\arraystretch{1.3}
\begin{tabular}{cc|ccc}
\multicolumn{2}{c|}{$Z_1(4050)$}&\multicolumn{3}{c}{$Z_2(4250)$} \\
 Fig.~\ref{fig:diag}(a) &Belle~\cite{belle_z4050}
      & \multicolumn{2}{c}{Fig.~\ref{fig:diag}(b)} &Belle~\cite{belle_z4050} \\\hline
 $1^-$ &  $?^?$ & $1^+$ & $1^-$  & $?^?$\\
 $4041 \pm 1$ &  $4051\pm 14^{+20}_{-41}$ & $4247 \pm 53$ &$4309 \pm 116$ &  $4248^{+44}_{-29}{}^{+180}_{-35}$\\
 $115 \pm 17$ & $82^{+21}_{-17}{}^{+47}_{-22}$ & $345 \pm 67$&$468 \pm 90$  & $177^{+54}_{-39}{}^{+316}_{-61}$\\
\end{tabular}
\end{ruledtabular}
\end{table}
We generate 
the Dalitz plot distributions for 
$\Lambda=1, 1.5$, and 2~GeV as in Fig.~\ref{fig:cutoff},
fit them as described above, 
and present the resulting ranges of the
Breit-Wigner parameters
in Table~\ref{tab:BW_param}
along with those of $Z_1(4050)$ and $Z_2(4250)$
from the Belle analysis~\cite{belle_z4050}.
The agreement is quite good for $Z_1(4050)$.
Meanwhile, the $Z_2(4250)$ mass and width from the Belle analysis
have rather large errors, and thus our results
for both $J^P=1^\pm$ assignments
easily agree with them.

\begin{figure}[t]
\begin{center}
\includegraphics[width=.5\textwidth]{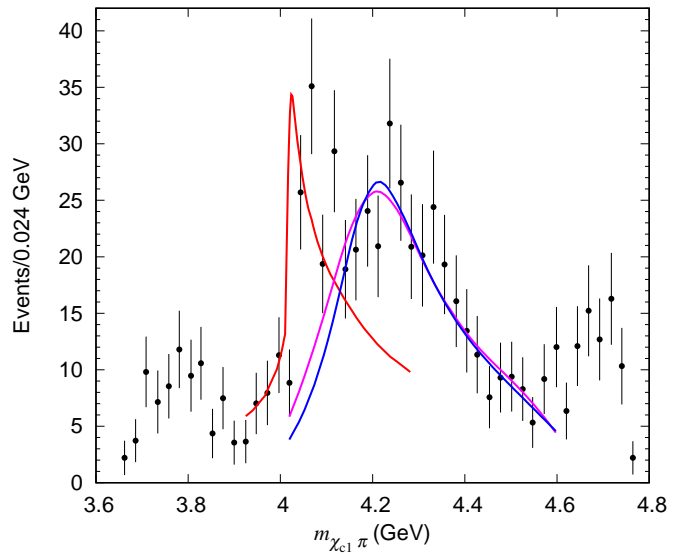}
\end{center}
\caption{
$\chi_{c1}\pi^+$ invariant mass distributions for 
 $\bar B^0\to\chi_{c1} K^-\pi^+$.
 The red, blue, and magenta solid curves in 
 Figs.~\ref{fig:spec}(a-c) are modified to include only
the contributions in 1.0~GeV$^2< m^2_{K^-\pi^+}<1.75$~GeV$^2$,
and superimposed on the Belle data
 (Fig.~14 of Ref.~\cite{belle_z4050})
 from the same kinematical constraint.
Each of the curves is multiplied by a constant factor
 and an incoherent constant background is added
 to fit the data.
 }
\label{fig:spec_data}
\end{figure}
\begin{figure}[t]
\begin{center}
\includegraphics[width=.46\textwidth]{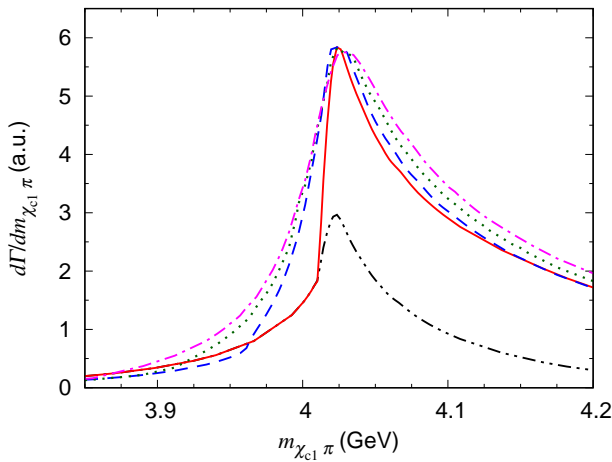}
\end{center}
 \caption{
$\chi_{c1}\pi^+$ invariant mass distributions for 
 $\bar B^0\to\chi_{c1} K^-\pi^+$ calculated with the triangle diagram A.
 The red solid, blue dashed, green dotted, and magenta dash-dotted curves
are calculated using 
 the $X(3872)\pi^+$ threshold energy smaller than the
 PDG value by 0, 50, 100, and 150~MeV, respectively;
see the text for details.
 All these curves, being scaled, have the same peak height.
The black dash-two-dotted curve is obtained from the red solid one by
 turning off the on-shell $X(3872)\pi^+$ contribution.
 }
\label{fig:spec_thres}
\end{figure}

Let us superimpose the spectra from the triangle diagrams A
and B on the Belle data (Fig.~14 of Ref.~\cite{belle_z4050}) as shown
in Fig.~\ref{fig:spec_data}.
Although this is a qualitative comparison where any interferences among
different mechanisms are not taken into account, 
the spectrum bumps from the triangle diagrams
capture characteristic features of the data.
In particular, the asymmetric shape from 
the triangle diagram A, which has a very sharp rise and a moderate
fall-off, is exactly what the data show.
In the Belle analysis~\cite{belle_z4050} where the Breit-Wigner form was
used to simulate this bump, their model does not seem to fit
this sharp peak of the data very well, as seen in Fig.~14 of the reference,
perhaps because the Breit-Wigner shape is not what the data call for.
As seen in Fig.~\ref{fig:spec}(a),
the spectrum shape from the triangle diagram A is 
significantly different from the Breit-Wigner.

It would be worthwhile to address
how this peculiar asymmetric shape comes about from the triangle diagram A.
By closely observing the spectrum shown 
in Fig.~\ref{fig:spec}(a) or an enlarged one shown 
by the red solid curve in
Fig.~\ref{fig:spec_thres}, the sharp rise of the spectrum starts from
an abrupt bend at $m_{\chi_{c1}\pi}\sim 4.01$~GeV where the $X(3872)\pi^+$
channel opens.
This implies that the sharp rise is assisted
by the opening of the $X(3872)\pi^+$ channel.
We indeed confirm this idea, as shown by the black dash-two-dotted curve
in Fig.~\ref{fig:spec_thres},
 by turning off the on-shell 
$X(3872)\pi^+$ contribution arising from $+i\epsilon$ in the denominator
of Eq.~(\ref{eq:amp}).

The proximity of the $X(3872)\pi^+$ threshold to the TS energy 
($\sim 4.025$~GeV) is also 
important to create the large asymmetry.
To see this, 
we change the $X(3872)$ and
$K^*(892)$ masses to lower the $X(3872)\pi^+$ threshold while
keeping the spectrum peak position almost the same.
We use, in unit of MeV, $(m_{X(3872)}, m_{K^*(892)})$=(3822, 1084),
(3772, 1218), and (3722, 1330) to lower the 
threshold by 50, 100, and 150 MeV, respectively.
The spectra calculated with these altered masses, presented in 
Fig.~\ref{fig:spec_thres}, show that
the rise of the bump
becomes significantly more moderate
as the threshold is lowered.
In this way, 
the $Z_1(4050)$-bump shape observed in
the Belle data is explained with well-founded physics, TS and
the channel opening near the TS energy, included in the triangle diagram A.

The asymmetric $Z_1(4050)$-bump shape could
sensitively discriminate different interpretations of $Z_1(4050)$.
A compelling model should explain not only the mass, width, and $J^P$
of $Z_1(4050)$,
but also its characteristic spectrum shape.
So far, only our model has successfully addressed this question.
It is also highly desirable to establish the spectrum shape 
with higher statistics data, 
considering that the Belle data still have large error bars.

It would be interesting to discuss the possibility of finding 
$Z_1(4050)$ and $Z_2(4250)$-like bumps in other processes. 
We point out that, actually, 
 the $Z_2(4250)$($J^P=1^+$)-like bump in $\bar{B}^0\to \chi_{c1}K^-\pi^+$
 and the $Z_c(4200)$-like bump in $\bar{B}^0\to J/\psi K^-\pi^+$~\cite{belle_z4200,lhcb_z4200}
 can be created by the same TS~\cite{ts_zc4430} from the triangle diagram B
 and thus are very similar.
Meanwhile, if the $Z_1(4050)$-like bump in
 the $\bar{B}^0\to \chi_{c1}K^-\pi^+$ data is generated by
the triangle diagram A, 
the same diagram but $\chi_{c1}$ replaced by $J/\psi$ should contribute
to $\bar{B}^0\to J/\psi K^-\pi^+$ because the $X(3872)\to J/\psi\pi^+\pi^-$
coupling is known to exist.
We calculated the $Z_1(4050)$-like spectrum for 
$\bar{B}^0\to J/\psi K^-\pi^+$
using the modified triangle diagram A; the pion-charmonium interaction
is now given by Eq.~(\ref{eq:contact2}).
The spectrum looks almost the same as the red solid curve of
Fig.~\ref{fig:spec}(a).
While a $Z_1(4050)$-like bump has not yet been observed in 
$\bar{B}^0\to J/\psi K^-\pi^+$~\cite{belle_z4200,lhcb_z4200},
the quality of the current data still leaves a possibility of finding it
in the $J/\psi\pi^+$ spectrum data of higher
statistics.
Although the possibility certainly depends on competitions with other
mechanisms, this is an interesting prediction from the TS-based
interpretation of the $Z_1(4050)$ bump.

\begin{figure}[t]
\begin{center}
\includegraphics[width=.5\textwidth]{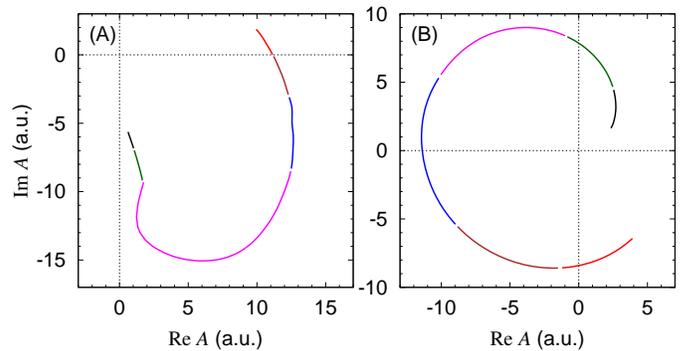}
\end{center}
\caption{
 Argand plots from the triangle diagrams A (left) and B (right),
corresponding to the spectra of 
Fig.~\ref{fig:spec}(a) and \ref{fig:spec}(b), respectively.
Six curved segments belong to six bins 
equally-separating the range of
 $M_{BW}-\Gamma_{BW}\le m_{\chi_{c1}\pi}\le M_{BW}+\Gamma_{BW}$; 
 $M_{BW}=4042$ (4194)~MeV and $\Gamma_{BW}=97$ (278)~MeV
 for triangle diagram A (B).
 $m_{\chi_{c1}\pi}$ increases counterclockwise.
 }
\label{fig:argand}
\end{figure}

Finally, we present Argand plots from the triangle diagrams A and B;
we use Eq.~(\ref{eq:contact2}) for the diagram B. 
Because $Z_1$ ($Z_2$) and $K^-$ are relatively in $p$-wave,
the angle-independent part of the amplitude is:
\begin{eqnarray}
A(m^2_{ab}) = \int d\Omega_{p_c}d\Omega_{p_{ab}}
 Y^*_{1,-s^z_{\chi_{c1}}}(\hat{p}_c)Y^*_{\ell 0}(\hat{p}_{ab}) M_{abc,H} \ ,
  \label{eq:argad}
\end{eqnarray}
with $\ell=0$ and 1 for the diagrams A and B, respectively;
$s^z_{\chi_{c1}}$ is the $z$-component of the $\chi_{c1}$ spin
and $m_{ab}$ the $ab$ invariant mass.
See Eq.~(B3) of Ref.~\cite{3pi} for the relation between 
the invariant amplitude $M_{abc,H}$ and 
$T_{abc,H}$ of Eq.~(\ref{eq:amp}).
$A(m^2_{ab})$ is
 shown in Fig.~\ref{fig:argand} as Argand plots. 
Both the triangle diagrams A and B create 
counterclockwise behaviors, seemingly similar to resonances. 

\section{conclusion}

We demonstrated that triangle singularities (TS)
from the triangle diagrams of
Figs.~\ref{fig:diag}(a) and \ref{fig:diag}(b) cause 
the bumps in the $\chi_{c1}\pi^+$ invariant mass distribution
of $\bar{B}^0\to \chi_{c1}K^-\pi^+$,
and that their positions and shapes, and thus Breit-Wigner parameters
fitted to the bumps, agree very well with those
found and named as $Z_1(4050)$ and $Z_2(4250)$
 in the Belle experiment~\cite{belle_z4050}.
 Within the resonance-based simulation of these bumps, 
 $J^P=1^-$ is predicted for $Z_1(4050)$ and 
 $J^P=1^+$ or $1^-$ for $Z_2(4250)$.
 The highly asymmetric shape of $Z_1(4050)$-like bump found by the Belle
 is well reproduced by our model; the opening of the $X(3872)\pi^+$
 channel near the TS energy causes the abrupt increase of the spectrum.
 This characteristic lineshape, which could discriminate different
 interpretations of $Z_1(4050)$,
 is yet to be accounted for by any other hadron
 structure models.
 We also discussed the possibility of finding bumps, caused by the same
 TS for $Z_1(4050)$ and $Z_2(4250)$, in
 $\bar{B}^0\to J/\psi K^-\pi^+$;
 $Z_c(4200)$ found in $\bar{B}^0\to J/\psi K^-\pi^+$ can be identified
 with $Z_2(4250)$ of $J^P=1^+$.
The kinematic effects,
TS and the channel opening, essentially
determine the shape and position of the spectrum bumps,
once the spin-parity of the $\chi_{c1}\pi^+$ system is specified by
Eqs.~(\ref{eq:contact})-(\ref{eq:contact3});
the uncertainty of the remaining dynamical details would not largely change the
presented results.

\begin{acknowledgments}
This work is in part supported by 
National Natural Science Foundation of China (NSFC) under contracts 11625523.
\end{acknowledgments}


\begin{thebibliography}{}

\bibitem{pdg}
M. Tanabashi et al. (Particle Data Group), Phys. Rev. D {\bf 98}, 030001 (2018).

\bibitem{belle_z4050}
R. Mizuk et al. (Belle Collaboration), Phys. Rev. D {\bf 78}, 072004 (2008).

\bibitem{babar_z4050}
J.P. Lees et al. (BaBar Collaboration), Phys. Rev. D {\bf 85}, 052003 (2012).

\bibitem{z4050_diquark1}
D. Ebert, R.N. Faustov, and V.O. Galkin,
Eur. Phys. J. C {\bf 58}, 399 (2008).

 \bibitem{z4050_diquark2}
S. Patel, M. Shah, and P.C. Vinodkumar,
Eur. Phys. J. A {\bf 50}, 131 (2014).

\bibitem{z4050_diquark3}
C. Deng, J. Ping, H. Huang, and F. Wang,
Phys. Rev. D {\bf 92}, 034027 (2015).

\bibitem{z4050_qcdsr1}
Z.-G. Wang, Commun. Theor. Phys. {\bf 63}, 466 (2015).

\bibitem{z4050_qcdsr2}
Z.-G. Wang, Eur. Phys. J. C {\bf 62}, 375 (2009).


\bibitem{z4050_molecule1}
X. Liu, Z.-G. Luo, Y.-R. Liu, and S.-L. Zhu,
Eur. Phys. J. C {\bf 61}, 411 (2009).

 \bibitem{z4050_molecule2}
 Y.-R. Liu and Z.-Y. Zhang, Phys. Rev. C {\bf 80}, 015208 (2009).

 \bibitem{z4050_molecule3}
 G.-J. Ding, Phys. Rev. D {\bf 79}, 014001 (2009).

\bibitem{z4050_qcdsr4}
S.H. Lee, K. Morita, and M. Nielsen,
Phys. Rev. D {\bf 78}, 076001 (2008).

 \bibitem{review_chen}
H.-X. Chen, W. Chen, X. Liu, and S.-L. Zhu,
Phys. Rep. {\bf 639}, 1 (2016).
	 
 \bibitem{review_raphael}
R.M. Albuquerque, J.M. Dias, K.P. Khemchandani, A. Martinez Torres,
	 F.S. Navarra, M. Nielsen, and C.M. Zanetti,
	 arXiv:1812.08207 [hep-ph].

\bibitem{landau}
L.D. Landau, Nucl. Phys. {\bf 13}, 181 (1959).

\bibitem{coleman}
S. Coleman and R.E. Norton,
Nuovo Cim. {\bf 38}, 438 (1965).

 \bibitem{s-matrix}
R. J. Eden, P. V. Landshoff, D. I. Olive and J. C. Polkinghorne,
The Analytic S-Matrix,
(Cambridge University Press, Cambridge, England, 1966).

 \bibitem{TS-Pc2}
M. Bayar, F. Aceti, F.-K. Guo, and E. Oset,
Phys. Rev. D {\bf 94}, 074039 (2016).

\bibitem{wu1}
J.-J. Wu, X.-H. Liu, Q. Zhao, and B.-S. Zou, Phys. Rev. Lett. {\bf 108}, 081803 (2012).
	
\bibitem{wu2}
F. Aceti, W.H. Liang, E. Oset, J.J. Wu, and B.S. Zou, Phys.Rev. D {\bf 86}, 114007 (2012).

\bibitem{TS-Pc}
F.-K. Guo, U.-G. Mei{\ss}ner, W. Wang, and Z. Yang,
Phys. Rev. D {\bf 92}, 071502 (2015).
	
  \bibitem{TS-Pc3}
X.-H. Liu, Q. Wang, and Q. Zhao,
	  Phys. Lett. {\bf B757}, 231 (2016).

  \bibitem{TS-a1}
M. Mikhasenko, B. Ketzer, and A. Sarantsev,
Phys. Rev. D {\bf 91}, 094015 (2015).

  \bibitem{TS-a1-2}
	  F. Aceti, L.R. Dai, and E. Oset,
Phys. Rev. D {\bf 94}, 096015 (2016).

 \bibitem{ts1}
S. Sakai, E. Oset and W.H. Liang,
Phys. Rev. D {\bf 96}, 074025 (2017).

\bibitem{ts2}
X.H. Liu and U.-G. Mei{\ss}ner,
Eur. Phys. J. C {\bf 77}, 816 (2017).

\bibitem{ts3}
S. Sakai, E. Oset and A. Ramos,
Eur. Phys. J. A {\bf 54}, 10 (2018).

\bibitem{ts4}
L.R. Dai, R. Pavao, S. Sakai and E. Oset, Phys. Rev. D {\bf 97}, 116004 (2018).

\bibitem{ts5}
Z. Cao and Q. Zhao,
Phys. Rev. D {\bf 99}, 014016 (2019).
	
\bibitem{ts6}
J.J. Xie and F.K. Guo,
Phys. Lett. B {\bf 774}, 108 (2017).

\bibitem{ts_zc4430}
S.X. Nakamura and K. Tsushima,
arXiv:1901.07385.
	
\bibitem{belle_z4430_2008}
S.K. Choi et al. (Belle Collaboration),
Phys. Rev. Lett. {\bf 100}, 142001 (2008).

 \bibitem{belle_z4430}
K. Chilikin et al. (Belle Collaboration),
Phys. Rev. D {\bf 88}, 074026 (2013).
	
\bibitem{lhcb_z4430}
R. Aaij et al. (LHCb Collaboration),
Phys. Rev. Lett. {\bf 112}, 222002 (2014).

\bibitem{belle_z4200}
K. Chilikin et al. (Belle Collaboration),
Phys. Rev. D {\bf 90}, 112009 (2014).

\bibitem{lhcb_z4200_Lb}
R. Aaij et al. (LHCb Collaboration), 
Phys. Rev. Lett. {\bf 117}, 082003 (2016).	

\bibitem{x3872_decay}
V. Bhardwaj et al. (Belle Collaboration), 
Phys. Rev. D {\bf 93}, 052016 (2016).

\bibitem{3pi}
H. Kamano, S.X. Nakamura, T.-S.H. Lee, and T. Sato,
Phys. Rev. D {\bf 84}, 114019 (2011).

\bibitem{lhcb_z4200}
R. Aaij et al. (LHCb collaboration),
Phys. Rev. Lett. {\bf 122}, 152002 (2019).
\end{thebibliography}


\end{document}